\documentclass[10pt,journal]{IEEEtran}
\usepackage{cite}

\usepackage[utf8]{inputenc} 
\usepackage[T1]{fontenc}
\usepackage[euler]{textgreek}

\usepackage{enumitem}

\usepackage{microtype}  
\usepackage{booktabs} 

\usepackage[top=2cm, bottom=2cm, left=3cm, right=3cm]{geometry}

\usepackage{url} 
\usepackage{lmodern} 

\usepackage{graphicx,dblfloatfix}
\usepackage{float}
\usepackage{subcaption}

\usepackage{wrapfig} 
\usepackage{amsmath}

\usepackage{algorithm}
\usepackage{algpseudocode}

\usepackage{epstopdf}

\begin{document}

\title{Prediction of Traffic Flow via Connected Vehicles }

\author{Ranwa~Al Mallah, Alejandro~Quintero, and~Bilal~Farooq
\thanks{R. Al Mallah and A. Quintero are with the Department of Computer Science, École Polytechnique de Montréal, CANADA, e-mails: ranwa.al-mallah@polymtl.ca and alejandro.quintero@polymtl.ca.}
\thanks{B. Farooq is with the Laboratory of Innovations in Transportation, Ryerson University, Toronto, CANADA, email: bilal.farooq@ryerson.ca. }}

\maketitle

\begin{abstract}
We propose a Short-term Traffic flow Prediction (STP) framework so that transportation authorities take early actions to control flow and prevent congestion. We anticipate flow at future time frames on a target road segment based on historical flow data and innovative features such as real time feeds and trajectory data provided by Connected Vehicles (CV) technology. To cope with the fact that existing approaches do not adapt to variation in traffic, we show how this novel approach allows advanced modelling by integrating into the forecasting of flow, the impact of the various events that CV realistically encountered on segments along their trajectory. We solve the STP problem with a Deep Neural Networks (DNN) in a multitask learning setting augmented by input from CV. Results show that our approach, namely MTL-CV, with an average Root-Mean-Square Error (RMSE) of 0.052, outperforms state-of-the-art ARIMA time series (RMSE of 0.255) and baseline classifiers (RMSE of 0.122). Compared to single task learning with Artificial Neural Network (ANN), ANN had a lower performance, 0.113 for RMSE, than MTL-CV. MTL-CV learned historical similarities between segments, in contrast to using direct historical trends in the measure, because trends may not exist in the measure but do in the similarities.\newline
\end{abstract}


\section{Introduction}

\IEEEPARstart{R}{oad} traffic congestion is a particular state of mobility where travel times increase. There is constant pressure on the authorities to take actions to improve the network traffic flow. To this end, predictive techniques are needed by infrastructure operators to allow advanced modelling. The real-time prediction of traffic flow on a road segment allows transportation authorities to take actions to control traffic load and prevent congestion \cite{feng2018adaptive, ryu2018construction}. Particularly, Short-Term traffic Prediction (STP) enables traffic managers to take informed decisions about how to best reroute traffic, change lane priorities and modify traffic light timing.

In the context of STP, many studies consider highways rather than urban regions \cite{pan2013short}. In a highway scenario, the road section can be represented as a network flow model that requires flow conservation on all segment. However, in an urban scenario, flow passing through the arc depend on multiple dynamic aspects that are difficult to describe and to model in detail \cite{djahel2015communications}. Urban networks are intricate complex networks so designing scalable traffic flow prediction models for these networks is required as they are more likely to be monitored by traffic managers.

Several unresolved problems exist for STP in urban networks. A traffic management system must firstly ensure efficient monitoring of the urban network. However, traffic state cannot be directly measured everywhere on the traffic road network because infrastructure operators are strained to monitor traffic while using the least possible resources \cite{abadi2015traffic}. Current collection methods rely on dedicated traditional heterogeneous sensor and backbone networks and hardware/software solutions. Due to the high complexity and uncertainty of contemporary transportation systems, these methods fail to capture in detail and in real time all the dynamics. Another problem with current traffic flow prediction models is their inadaptability of detecting and tracking the traffic patterns changes \cite{hou2015traffic}. There is a new pattern every time a non recurrent congestion occurs in the traffic flow and in this case, the model is not able to predict as accurately as when there is recurrent congestion. Currently, operators rely on external data sources to assess in real-time the different events happening on the road, such as special events, incidents and inclement weather. Future systems should enable continuous monitoring of the traffic condition along all roads of the traffic network based on real-time information because well-tailored data sources may not always be available for a particular area of the traffic network.  

With the advances in computing technology, such as Connected Vehicles (CV) technology enabling Vehicle-to-Infrastructure (V2I) and Vehicle-to-Vehicle (V2V) communications, transportation management is no more uniquely a transportation engineering problem. In fact, connected vehicles evolve in a data-rich environment where they consistently generate and receive a variety of data. In this article we show how integrating the transportation system with real-time information from connected vehicles for short-term traffic flow prediction on a target road segment results in a powerful tool for transportation analysis and evaluation. Self-organization is essentially a distinctive characteristic of CV and represents a new approach in this domain. 

We foresee that an accurate prediction requires a mixture of distributed and centralized architecture through leverage of vehicular communication. Via this next generation sensing technology, we are interested in identifying road traffic events on the basis of exchanging traffic flow data between vehicles. If CV can detect congestion and cooperatively attribute a possible cause to it, we believe that they can then transfer this knowledge in real time to a central entity able to accurately predict flow on a road segment. On the other hand, because the flow fluctuates from one time to another, it's better for a road side unit (RSU) to monitor the parameters for a period of time. 

Collecting microscopic and macroscopic traffic data from CV for traffic prediction purposes has already been proposed in the literature \cite{gramaglia2014abeona,leontiadis2011effectiveness}. Typically, at the low level, vehicles perform local, real-time sensing and previous studies extract information from the data collected. Unlike previous studies, we aim at extracting knowledge from the information and using the knowledge acquired for the sake of traffic prediction. We propose a Deep Neural Network (DNN), and tackle the problem by learning the target DNN in a multitask learning technique. DNNs have successfully been applied to traffic flow prediction \cite{vlahogianni2005optimized}. Their good performance lies in their capacity to embody the transportation network's large amount of data and high dimensions of features. When compared with state-of-the-art ARIMA time series and baseline classifiers such as Random Forest (RF), results show that our approach, namely MTL-CV, presents an average performance in terms of root-mean-square error (RMSE) equal to 0.05. Compared to single task learning with Artificial Neural Networks (ANN), our experiments show that ANN have a lower performance (0.113 for RMSE) than MTL-CV, but higher performances than ARIMA. This shows that when the tasks involved in multitask are semantically connected a larger improvement in accuracy of prediction can be obtained. Finally, we provide an analysis to show the impact of overestimating the performance of the wireless network. The challenges due to the communication network in terms of channel conditions, packet losses, collisions and delay have an impact on the traffic flow prediction. The estimation of the travel time index had an increase in error of 0.119. The performance of the model is even lower for flow values estimated by the CV, with an increase in error of 0.197. 

The contributions of this paper are summarised as follows:

\begin{itemize}

\item Monitoring of microscopic and macroscopic traffic variables via connected vehicles for the extraction of relevant contextual traffic features in order to summarise valuable knowledge in an efficient way.

\item Forecasting of short term traffic flow on a target road segment with a Deep Neural Network trained to predict multi tasks with input from connected vehicles. 

\item Evaluation and validation of the proposed framework and inference method is made relying on simulation generated scenarios from a realistic data set of urban city vehicular motion traces. 

\end{itemize}

This paper is organized as follows. After introducing the related work in Section II, we describe our proposed framework in Section III. In Section IV, we present the STP model. In Section V, we describe the simulation based on real vehicular mobility traces and provide analysis and discussion of the results. We conclude the paper in Section VI.

\section{RELATED WORK}

The traffic flow prediction problem aims at evaluating anticipated traffic flow at future time frames on a target road segment. The main techniques used by parametric approaches to solve the short-term traffic flow prediction problem are time-series models, Kalman filtering \cite{pan2013short} and AutoRegressive Integrated Moving Average (ARIMA)-based models \cite{williams2003modeling}. In \cite{hamed1995short}, they applied ARIMA to predict traffic in urban roads. Kohonen-ARIMA (KARIMA) \cite{van1996combining}, vector autoregressive moving average (ARMA),  ARIMA with explanatory variables (ARIMAX) \cite{kongcharoen2013autoregressive}, seasonal ARIMA (SARIMA) \cite{tseng2002fuzzy} and space–time ARIMA \cite{hamilton1994time} were all proposed to improve performance of prediction. Statistical models make the assumption of stationarity of the underlying process. However, traffic flow has a stochastic and nonlinear nature, unfortunately, even an enhanced ARIMA cannot accurately predict flow in the presence of accidents \cite{kongcharoen2013autoregressive}. ARIMA cannot be used in the presence of events happening on the road because events cause sudden changes in the time series data and ARIMA is slow to react. Moreover, while time-series analysis models are probabilistic, they are ignorant of the underlying process that generates the data. 

Nonparametric regression is used in \cite{smith1994short}. In \cite{wu2012online}, they propose a boosting regression to forecast flow under abnormal traffic conditions. In \cite{castro2009online}, they use a support vector regression and a particle swarm to optimize the model's parameters. Among all of these techniques, neural-network-based forecasting had the best performance in terms of prediction accuracy. Literature shows promising results when using artificial neural network models as they are used as benchmarking methods for short-term traffic prediction \cite{liu2011discovering}. Typical methods include back propagation (BP) neural network, radial basis function (RBF) neural network, recurrent neural network and time-delayed neural network \cite{ karlaftis2011statistical}. 

Particularly, deep learning is a neural network of more than one hidden layer. This technique has attracted researchers from various domains as it considers complex correlations between features and outputs. In \cite{chen2017traffic}, they propose a stacked auto-encoder model to learn generic traffic flow features by considering the spatial and temporal correlations. Moreover, recent work has shown that it is possible to jointly train a DNN for solving different tasks simultaneously \cite{huang2014deep}, MultiTask Learning (MTL). By sharing what they learned, the model learns them together better than in isolation. In \cite{sun2012network} and \cite{huang2014deep}, they train a MTL model to predict flows on links. The authors propose multilink forecasting models, which take the relations between adjacent links into account. Since each link is closely related to other links in the whole transportation system, the multilink model predicts traffic flows using historical traffic flow data from all of the adjacent links. The features in \cite{sun2012network} are flow data collected from sensors on the road. In \cite{huang2014deep}, they propose a deep architecture that consists of two parts, i.e., a deep belief network (DBN) at the bottom and a multitask regression layer at the top. Since all roads are connected to each other, alot of shared information coexists. They collect data from inductive loops deployed on freeways.  In urban networks however, the impact of traffic is a complex multi-dimensional problem.

Because studies focus on traffic history and neglect other conditions affecting traffic, in \cite{koesdwiady2016improving}, they investigate and quantify the impact of weather on traffic prediction in a freeway scenario. In fact, weather conditions may have a drastic impact on travel time and traffic flow. Their MTL architecture incorporate deep belief networks for traffic flow prediction using weather conditions. Flow data is measured every 30s using inductive loops deployed on freeways. Other studies investigate hybrid approaches by combining several techniques\cite{vlahogianni2005optimized}. None of them incorporate events into the traffic flow prediction model, they predict flow in typical rush hour scenarios. More recently, some model were able to predict flow on a target road segment in the presence of one type of event, not in the presence of any event because it becomes more challenging when considering different causes of congestion in the prediction. 

A survey by \cite{djahel2015communications} reported that the design principle of a traffic flow prediction algorithm is to use a combination of historical data, real-time feeds, traffic modelling and simulation. Also, they inform that real-time monitoring of traffic should be done at short intervals to provide good quality because stale data is useless in dynamic environments. To the best of our knowledge, we are the first study to consider the combination of historical traffic patterns with the real-time traffic information collected by connected vehicles for the problem of traffic flow prediction in an urban road network. We see that a much more efficient system would result if the vehicles of the connected vehicles themselves collect real time feeds because computation would aggregate a quality of data at a vehicle level. 

The focus of our study is to integrate via connected vehicles the impact of various events into forecasting of short-term traffic flow. We consider spatiotemporal characteristics of traffic in training our models because no studies have tackled the problem of analysing the tight correlation between traffic data and external factors in an urban traffic network via the CV technology. It should be noted that the prediction of traffic flow under atypical conditions is evidently more challenging than doing so under typical conditions and, hence, much desired by operational agencies. 

\section{FRAMEWORK}

The framework proposed in this study is semi-centralized. On one hand, CV collect and propagate data via the ad hoc networks formed between them along a route. On another hand, a Road Side Unit (RSU) is installed on a target road segment and collects data for a period of time to get a clearer picture about the traffic on the target road segment where flow needs to be predicted. 

\subsection{Data collection by CV}

\begin{figure*}[!t]
 \begin{center}
 \includegraphics[width=0.5\linewidth,height=9cm]{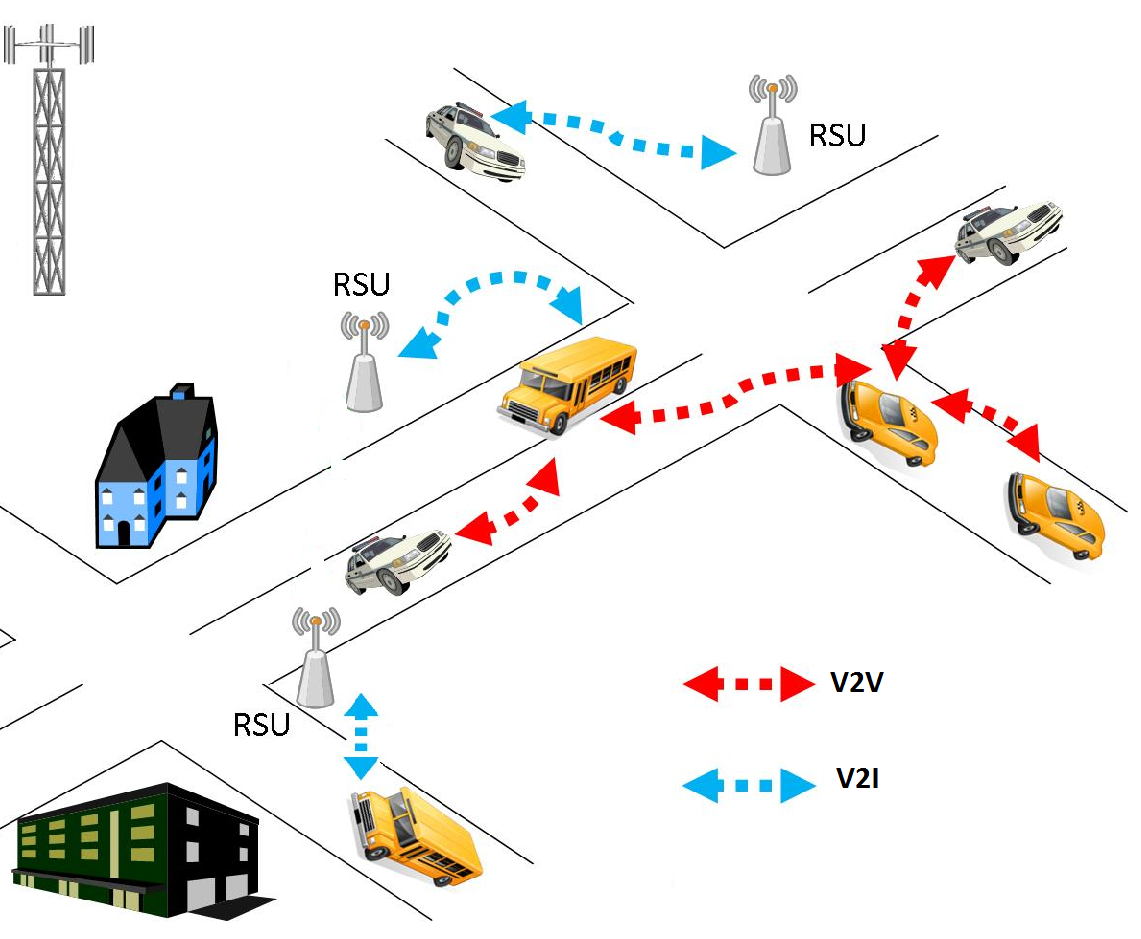}
 \caption{Propagation process via V2V and deployment of a RSU on the target segment}
 \label{fig:RSU_6}
 \end{center}
\end{figure*}

Each connected vehicle continuously collects traffic characteristics along every segment of its trajectory. Vehicles use broadcasting as a data forwarding, allowing data to move faster than the speed of traffic. In our design, vehicles are not required to be continuously connected to each other. In fact, a vehicle should be able to enter Store-Carry-Forward (SCF) mode if there are no vehicles in his vicinity. Upon investigation of data on board of the vehicle, each vehicle computes travel time on each segment of its trajectory. Travel time is the key data in this study. CV also collect travel times of others by cooperation between them. Then, each vehicle compares its assessment with those received from its surroundings so as to update it. The vehicle then broadcasts its traffic data. This propagation process is shown in Fig. \ref{fig:RSU_6}. We highlight that computing travel time from speed measurements would be an estimation of the travel time rather than the real travel time of the vehicle on the segment. In fact, in an urban road network, the speed of a vehicle varies every second according to other drivers behaviour, traffic light signals, the road, characteristics of the vehicle. Connected vehicles naturally extract the travel time experienced along a segment before entering another one. 

Each time a vehicle receives travel time information broadcasted by another vehicle, it updates its stored data accordingly. The vehicle firstly computes a TravelTime Index of their own, \textit{TTindex}, and secondly averages travel time indexes of others. The index is representative of the observed travel time \textit{TTi} of the vehicle on segment \textit{i} compared to the historical travel time value on each segment \textit{TThi} along the trajectory. The equation is as follows:

\begin{equation}					
TTindex= \frac{\sum_{i=1}^{10}(1-exp^{-\frac{i}{10}(\frac{TTi-TThi}{TThi})})}{10}
\end{equation}

We vary \textit{i} between 1 and 10. Precisely, the number '10' refers to the last 10 segments of a vehicle's trajectory. The index thus represents the sum of the weighted average of the difference in travel time on a link, for 10 previous segments of the trajectory. The weights in the index around the current segment increase so to better capture current local view. In fact, the larger the \textit{i}, the more the contribution to \textit{TTindex}. 

We impose the vehicles compute data from the last 10 segments of their trajectory to limit the analysis to a specific area. The environment of the study dictated the appropriate number of segments to be considered. In fact, the method provided in the study was applied to an urban road network containing links having different lengths. It is considered a 'complex network' that mimics the real-life context of vehicle mobility. In the urban network of our study, approximately 10 segments of signalized arterials corresponded to an area of 2 km. As suggested by transportation engineer experts, flow on a road segment is correlated to its surrounding and on average a node has an area of interest of 2 km. We assume that a traffic situation happening beyond 2 km of a target segment will not have an impact on the prediction of flow on a target segment. In general, the method in the study can permit vehicles to store information of more than 10 segments of their trajectory. In fact, the method can also be applied to highways or other cities by adapting the required collect effort (in terms of number of segments of the trajectory) depending on the environment. 

After computing the travel time index of their own, vehicles average travel time indexes of others. In order to reduce randomness, the average method is taken to calculate the travel time and considering the impact of the travel times of all vehicles within the scope, the formula is set as follows:
 
 \begin{equation}					
v= (1-\alpha)*v\char`_s + (\alpha)*v\char`_r
\end{equation}
 
where \textit{v\char`_s}  is selected as the travel time index of the vehicle, \textit{v\char`_r} is a mean value of the vehicular travel time indexes in the wireless coverage of the vehicle and $\alpha$ is a weighting factor representing the different degrees of importance. After experimentation we fixed $\alpha$ to be 0.65. 

Precisely, each vehicle reports a travel time index that represents the travel time around a segment of interest. Since every vehicle comes into the segment of interest from a different trajectory, the travel time indexes of all vehicles must represent the travel time on all the segments that are directly adjacent to the segment of interest. To estimate the value of $\alpha$, the experimentation went as follows:

- We ran simulations and assessed travel time information on the 10 segments that are adjacent to the segment of interest. We computed with equation (1) a ground truth  \textit{TTindex} representing the perfect case where one vehicle would have visited all 10 road segments at the same time.

- We introduced the weighting factor $\alpha$. If $\alpha$ equals 0 only the \textit{TTindex} computed by the connected vehicle on the target segment is taken into account in the calculation. If $\alpha$ equals 1, the average \textit{TTindex} of all other vehicles on the segment of interest is taken into account and not that of the vehicle computing the value. To get as close as possible to the ground truth value, we varied $\alpha$ in equation (2) and compared the resulting \textit{TTindex} value with that of the ground truth.

Results showed that on average, values of $\alpha$ ranging between 0.6 and 0.75 were representative of the ground truth \textit{TTindex}. This means that relying slightly more on other vehicle's assessment was more representative of the traffic situation around the segment. We fixed $\alpha$ to 0.65.

In the absence of vehicles around them, CV collect their data and compute a \textit{TTindex} without input from other vehicles. The \textit{TTindex} feature can be computed without paths being established if traffic is low. Vehicles are not required to be continuously connected to each other. SCF can be employed where the vehicles carry the data and travel to deliver it to another vehicle. \textit{TTindex} is computed on a per-segment basis. Every time a vehicle changes segment, the vehicle computes a \textit{TTindex} and incorporates values of \textit{TTindex} of surrounding vehicles (if there are any) in the calculation. If connections can be established between vehicles, even if they are short-lived or even if the data is sent from a vehicle who employed a SCF technique to deliver the information, vehicles can exchange TTindex values as this feature represents by itself the freshness of the travel time data in space and time. For the period vehicles remain in the coverage range of each other, they will continue to receive \textit{TTindex} values from each other and should update the values in their tables so as to compute a fresher TTindex of their own.

Finally, the framework incorporates the ability of registering in vehicle detailed information on the transient altering events along a vehicle's trajectory. To do so, we implement on board of each vehicle the algorithm for the detection of congestion via connected vehicles presented in \cite{younes2015performance}. Also, we implement the algorithm in \cite{al2017distributed} that permits vehicles classify the cause of the detected congestion.  The results of the local real-time monitoring done by the CV is disseminated reactively to others on the segment. 

\subsection{Data collection by RSU}
On the target road segment where traffic flow prediction is required, an RSU is installed and continuously stores \textit{TTindex} values from vehicles passing on the segment. Vehicles on neighbouring segments, in the coverage range of the RSU or not, are not required to relay \textit{TTindex} values to the RSU as this information is not taken into consideration in the forecasting. There is no need for a vehicle to establish a path from outside of the range of the RSU to relay \textit{TTindex} values to the RSU. In order to predict flow on a target segment, an RSU can have more knowledge by receiving and gathering features originating first and foremost from the vehicles on the target segment. CV on that segment are involved in the process and data collection by the RSU will always be done only within its coverage range. 

\textit{TTindex} is a spatio-temporal feature. The weights in the index around the current segment increase so to better capture current local view. Under such circumstances, \textit{TTindex} data accounts for spatial and temporal relevance and is computed in such a way that it will always remain valid at the time of its delivery to the RSU. 

Also, the collect effort would have to be restricted in terms of duration. To this end, each vehicle creates and stores traffic characteristics in its database. The database consists of the vehicle's own measurements and measurements received from other vehicles with a timestamp. The time of the measurement's creation ensures freshness of measurements and helps prioritize most recent ones. Maintaining the database consists in deleting measurements that are older than a time threshold. 

The time threshold that determines when a measurement created by the vehicle itself or received by others becomes outdated is 15 minutes.  Measurements and events that are older than 15 minutes are not reported to the RSU. This limits considerably the quantity of information to be stored at a vehicle level. The spatio-temporal restriction ensures freshness of the information and accounts for the finite vehicle buffer capacity.

Also, each vehicle sends the \textit{vehicleID}, \textit{time}, \textit{TTindex}, \textit{Segmentid10}, \textit{Flow10}, \textit{event10}, ..., down to \textit{Segmentid1}, representing the last ten segments of their trajectory. If for some reason the communication network is congested and the vehicle is not able to upload all its data to the RSU, the latter would have information on a subset of segments only. Another vehicle would have transmitted information on another subset of segments of interest to the RSU so as to complete the RSU's vision of influenced road segments. Also, uploading information on the latest segments of a vehicle trajectory orderly represents a strategy of selective uploading. 

Finally, the RSU collects information and does the mapping between all influenced road segments. Around the target segment, the influenced road segments are those whose flow readings vary a lot compared to expected flow when an event occurs in the surrounding. 

\subsection{Implemented algorithms}

We summarize in Fig. \ref{fig:Procedures}  the algorithms implemented on board of each vehicle and at the RSU. 

\begin{figure*}
\textbf{Algorithm} - \textit{$V_i$}: A vehicle in the scenario, \textit{oTT}: Observed or current travel time of \textit{$V_i$}, \textit{TTh}: Historical or expected travel time of \textit{$V_i$}, \textit{CurrentTT($V_i$)}: Travel time of the vehicle on the road and local traffic evaluation by testing if travel time is excessive and congestion is detected on the segment, \textit{EdgesOfRouteofV}: Table to store information from BEACON messages, \textit{ListEdges}: Table to store computed information of the last ten segments of the trajectory, \textit{ADJ-SegmentID}: an adjacent segment to the target segment the RSU is on. \\
 
\begin{algorithmic}[1]
\If{Connected Vehicle}
\State \textit{$V_i$} broadcasts BEACON message to its neighbors
\State Get current road segment of \textit{$V_i$}	
\State \textit{$V_i$} receives BEACON  messages		
\State Compute CurrentTT(\textit{$V_i$}) and flow on the segment for local traffic evaluation	
\State Compute own TTindex
\State Compute TTindexAverage with collected TTindex of others in table EdgesOfRouteofV 
\State Broadcast TTindexAverage
\State Compute variables as per [ …]
\If{\textit{oTT} > 1.8 * \textit{TTh}}		// ANALYSIS 
\State Estimate the cause of the excessive congestion and store it as an event on the edge in ListEdges.
\EndIf
\EndIf
\If{RSU}	
\State RSU collects BEACON messages in a table
\State Computes flow regularly on target segment and stores it in HistoricalFlowTable 
\State Computes average TTindex from all BEACON received
\State Receives from vehicles the content of their ListEdges 
\State Compares SegmentID received to all ADJ-SegmentID
\State Extracts flow and event only if applicable 
\State Generate a feature vector 
\State Predict future flows
\EndIf
\end{algorithmic}
\caption{Data collection done by CV and the RSU }
\label{fig:Procedures}
\end{figure*}

\subsubsection{Algorithm implemented on board of each vehicle}
Procedures on lines 2-4 consist of the monitoring phase. Each connected vehicle \textit{$V_i$} broadcasts every 0.1 seconds a BEACON message to its neighbours. \textit{$V_i$} also receives BEACON messages from others. Knowing its current road segment, BEACON messages received from other segments are dropped. The connected vehicle stores the traffic characteristics received in an information structure, EdgesofRouteofV, containing the following fields: \textit{SegmentID}, \textit{Time}, \textit{SenderID}, \textit{Position}, \textit{Speed}, \textit{TTindex} and other fields. If a BEACON message is received from a \textit{SenderID} already in the list, \textit{$V_i$} updates the stored information. Measurements that are older than 15 minutes are automatically discarded.

Procedures of lines 5-8 represent the aggregation phase. \textit{$V_i$} computes flow on the segment for local traffic evaluation and stores it in ListEdges. This list contains the following fields: \textit{SegmentID}, \textit{Time}, \textit{CurrentTT}, \textit{Event} and \textit{Flow} on the segments of the vehicle's trajectory. If the list has less than 10 elements, the new entry is added. Otherwise, if the list has 10 entries, the last one in the list is deleted and the new entry corresponding to the current road segment is added on top. This ensures the list has utmost 10 entries. From the current travel time computed by the connected vehicle on the segment and the travel times of previous edges stored in ListEdges, \textit{$V_i$} computes its own \textit{TTindex} value. With equation (2), \textit{$V_i$} averages its \textit{TTindex} with travel time indexes of others in table EdgesofRouteOfV. 

Procedures of lines 9-12 represent the analysis phase. We compute variables as per \cite{al2017distributed}. If the observed travel time is above a threshold, the vehicle estimates the cause of congestion by creating the feature vector and inferring with the classifier the cause. The vehicle then stores the cause as an event on the edge in ListEdges.

\begin{figure*}[!t]
\centering
\includegraphics[height=185px, width=400px]{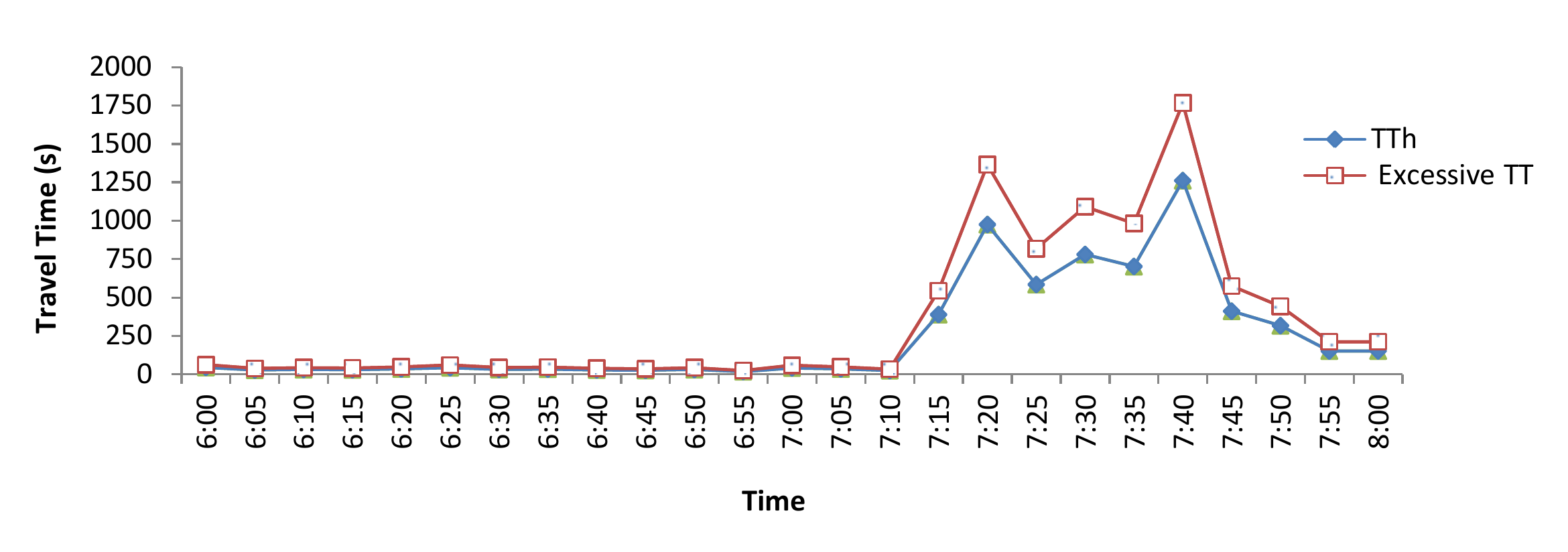}
 \caption{Historical \textit{TT} on a segment at 5 minutes interval and excessive \textit{TT} with c=1.4}
 \label{fig:TTh}
\end{figure*}

Precisely, if the observed travel time is higher than a threshold, which is determined by multiplying the congestion factor \textit{c} with the expected recurring delay on the particular segment, the travel time is said to be excessive, or else, it is considered normal. Fig. \ref{fig:TTh} shows the expected recurring travel time \textit{TTh} on some segment of the road traffic network. Expected travel times can be extracted from history based on the fact that the average speed on roads is usually similar at the same time of different days of the week. In a real deployment, expected values can be derived offline using past historical data for each segment. The preloaded digital maps available on the vehicular nodes may provide this traffic statistic of the roads at different time of the day.

\begin{figure*}[!t]
\centering
\includegraphics[height=150px, width=450px]{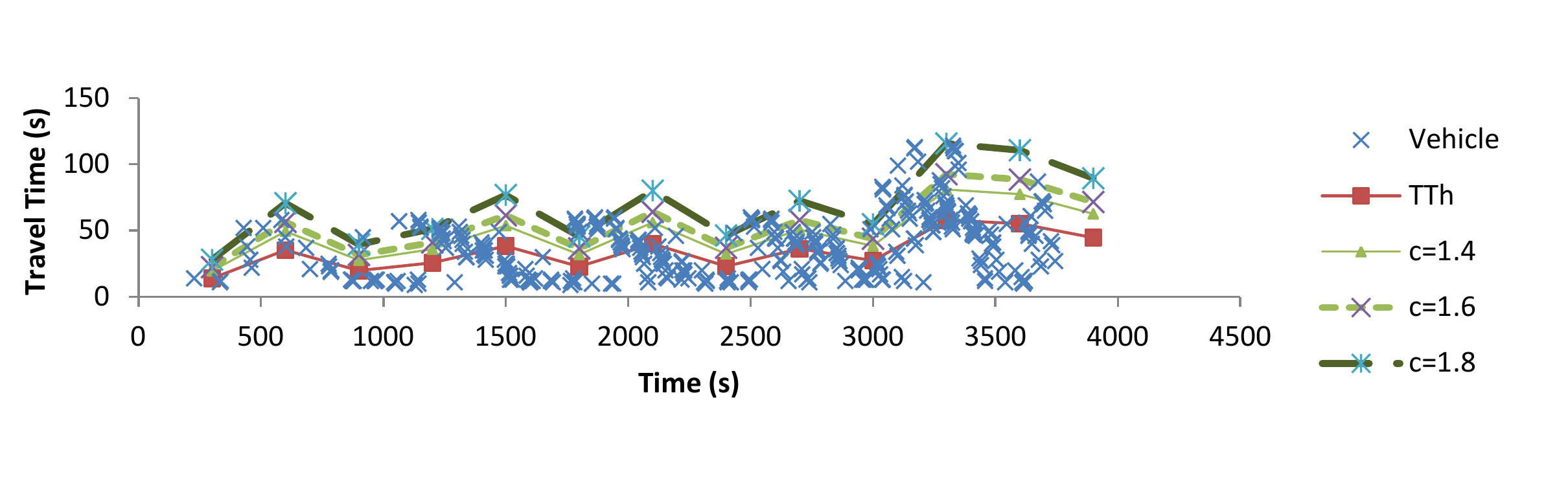}
 \caption{Variability of travel time data}
 \label{fig:variability}
\end{figure*}

Particularly, we applied the same technique as in \cite{anbarouglu2015non} to examine the travel time variability and get the appropriate value of \textit{c} for our scenario. Other cities may have different optimal values for \textit{c} and this has been thoroughly studied by the authors in \cite{anbaroglu2014spatio}. Fig. \ref{fig:variability} shows the variability of the collected values of travel times for vehicles on a particular segment. We studied different road segments with an average of 800 vehicle data points per segment. We varied the congestion factor to find the appropriate value of \textit{c} permitting vehicles not to assess excessive travel time in absence of non-recurrent congestion. For most segments in the base scenario, when \textit{c} = 1.8, most data points were below the curve, this is shown in Fig. \ref{fig:variability}. We fix \textit{c} = 1.8 for the urban network in our study.

\subsubsection{Algorithm implemented at the RSU}

The infrastructure also receives messages from vehicles. Knowing the road segment it is on, messages received from other segments are dropped. RSU stores the traffic characteristics received in an information structure containing the following fields: \textit{Time}, \textit{SenderID}, \textit{Position}, \textit{Speed}, \textit{TTindex} and \textit{Direction}. If a message is received from a \textit{SenderID} already in the list, RSU updates the stored information. Measurements that are older than 15 minutes are automatically discarded. RSU regularly computes flow on the target segment and stores the values in HistoricalFlowTable, a table containing only the past four traffic flow values calculated. 

When the RSU is ready to predict traffic flow on a segment, it computes the average \textit{TTindex} from all messages received. On one hand, in the framework, the RSU has a list of eight adjacent segments, ADJSegmentID, that affect flow on the target segment. On the other hand, vehicles start sending to the RSU the content of their ListEdges (containing fields: \textit{SegmentID}, \textit{Time}, \textit{CurrentTT}, \textit{Event} and \textit{Flow} on the last ten segments of their trajectory). If the value of \textit{SegmentID} received by the RSU from the vehicle is equal to one of the eight ADJSegmentID values, the RSU extracts from the message the flow and event on the segment as reported by the vehicle. Otherwise, the RSU drops the message received by the vehicle. The RSU generates a feature vector (time, currents flows/events on adjacent segments, \textit{TTindex}, past traffic flows) in order to predict future flow.

Feature selection is one of the core concepts in machine learning and has a huge impact on the performance of the model. We believe we identified the features that contribute the most to our target variable in order to achieve the best performance. The purpose is to use the features to predict traffic flow by means of a deep learning technique that learned from an adequate dataset to automatically infer from the correlations between these variables. The predictor takes input of the features sent to the RSU, and output the predicted flow. The predictor is the STP model and is presented in the next section. 

\section{STP Model}

The traffic flow prediction problem can be stated as follows. Let \textit{$X_i$(t)} denote the observed traffic flow during the \textit{t}th time interval at the \textit{i}th segment in a transportation network. Given a sequence of observed traffic flow data, \textit{i = 1, 2, . . . , m}, and \textit{t = 1, 2, . . . , T} , the problem is to predict the traffic flow at time interval (t+$\Delta$) for some prediction horizon $\Delta$.  

On the other hand, most models in the literature predict flow \textit{$X_i$(t+$\Delta$)} at time\textit{ (t+$\Delta$)} based on the traffic flow sequence \textit{X = \{$X_i$, t|i $\epsilon$ O, t = 1, 2, . . . , T \}} in the past, where \textit{O} is the full set of observation points. The problem becomes, given the feature \textit{X} and task \textit{Y} pairs obtained from history traffic flow \textit{\{($X_1$, $Y_1$), ($X_2$, $Y_2$), . . . , ($X_n$, $Y_n$)\}}, learn the best parameters for a prediction model that minimizes a loss function. This is supervised learning because each input can be tagged with the flow \textit{Y} corresponding to the next value in the time series obtained offline. 

\begin{figure*}[!t]
\begin{center}
\includegraphics[width=0.9\linewidth,height=10cm]{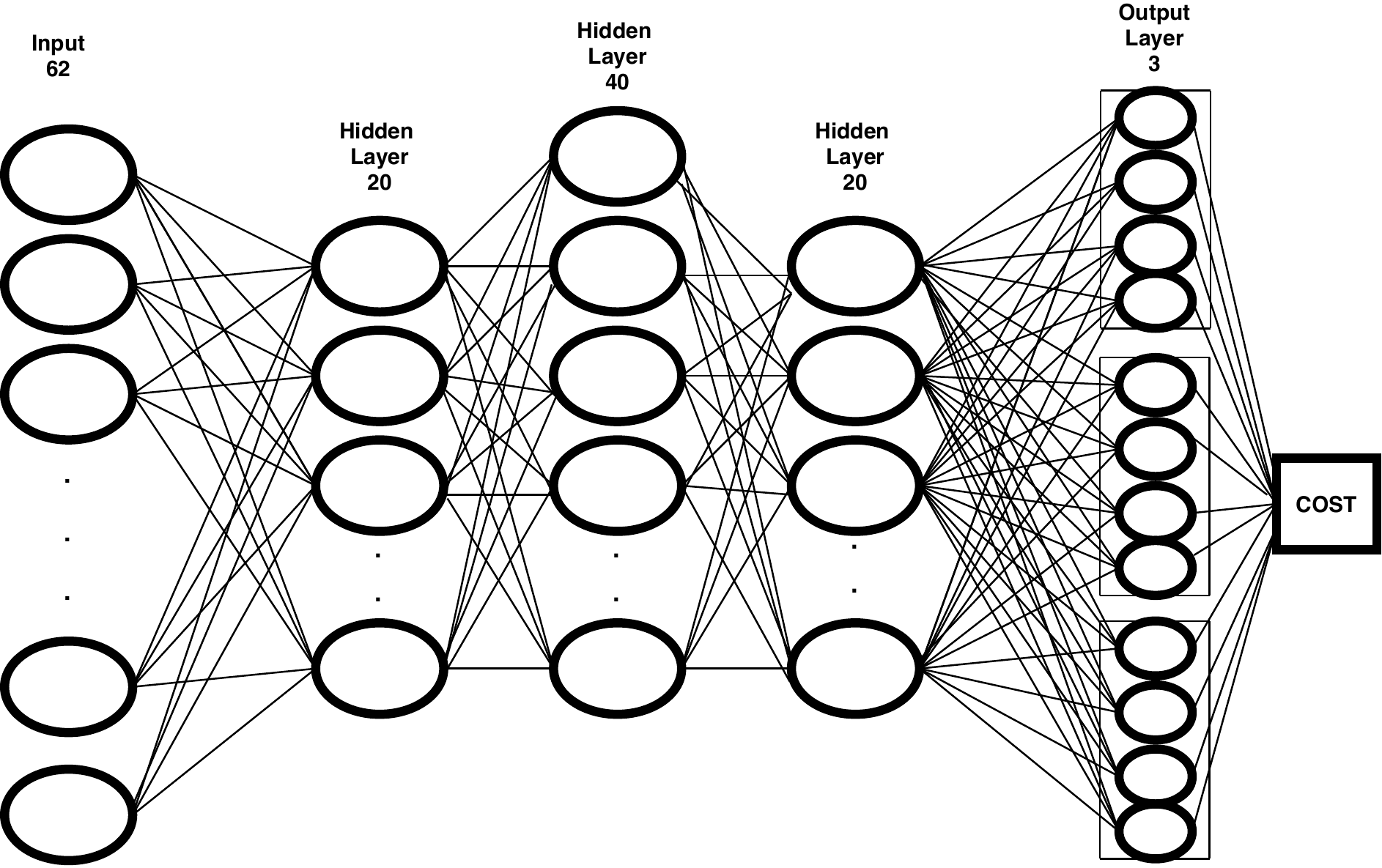}
\caption{Multitask learning DNN}
\label{fig:model}
\end{center}
\end{figure*}

However, in our work, we incorporate to the input feature \textit{X}, not only previous traffic flows observed on the target road segment but knowledge acquired from related roads. The features are: current time of day, observed travel time trajectory of vehicles around the target segment (TTindex), past successive flow values on the target segment, flows on links around the segment and the presence of any traffic event on surrounding segments, such as incident, weather, special event, workzone or recurrent traffic. 

Particularly, in this study, the problem of predicting short-term flow is handled as a classification task. In fact, we propose that the target variable \textit{Y} represent multiple classes of discrete interval of flows and the task is for the classifier to predict the range of flow that the current traffic situation will generate at a near future time. Moreover, we propose that the classifier learns to solve multiple tasks at the same time. Precisely, we use a single model that is able to do MTL via multiple outputs, each corresponding to the same task at a three different time. The output used for short-term flow prediction would be the middle one so that there are tasks earlier and later that the model trained on. In particular, we propose that given a fresh new road network traffic situation at time \textit{t}, the first task, \textit{$X_t+5$}, consists in determining what flow \textit{k} $\epsilon$ \textit{Y} is a suitable flow prediction at \textit{t+5}. The second task is to find what flow \textit{k} $\epsilon$ \textit{Y} is a suitable short-term flow prediction at \textit{t+15} based on the similar road network traffic situation and on the relevant prediction of the first task and the third task is to find the flow at \textit{t+20}. 

We propose a Multi-Layer Perceptron (MLP) that solves the three tasks. The MLP is a series of logistic regression models stacked on top of each other. The hidden units learn non-linear combinations of the original inputs. The last layer is also a logistic regression because we are solving a classification problem. Deep learning is a promising approach for transportation networks because they are highly correlated and they generate a large amount of data with high dimensions of features. Thus, we extend the MLP to a Deep Neural Network, and tackle the problem by learning the target DNN in a multitask learning technique. The supervised multitask learning DNN model is presented in Fig. \ref{fig:model}. 

The input of the joint learning model is of dimension 62 and feeds three hidden layers. The input vector consists of the features combined from CV and RSU. The supervised classifier has 20, 40 and 20 hidden units in the different layers. All hidden layers are fully connected. Three outputs are fully connected to the last hidden layers. Each output of the network contains four neurons representing the class label. The multitask model is trained for classification on labelled examples. The target variable represents whether a traffic situation will generate one of four ranges of flow. The output is passed to three independent softmax to produce the scores for the individual tasks. 

\begin{figure}[H]
\begin{center}
\includegraphics[width=1\linewidth,height=7.5cm]{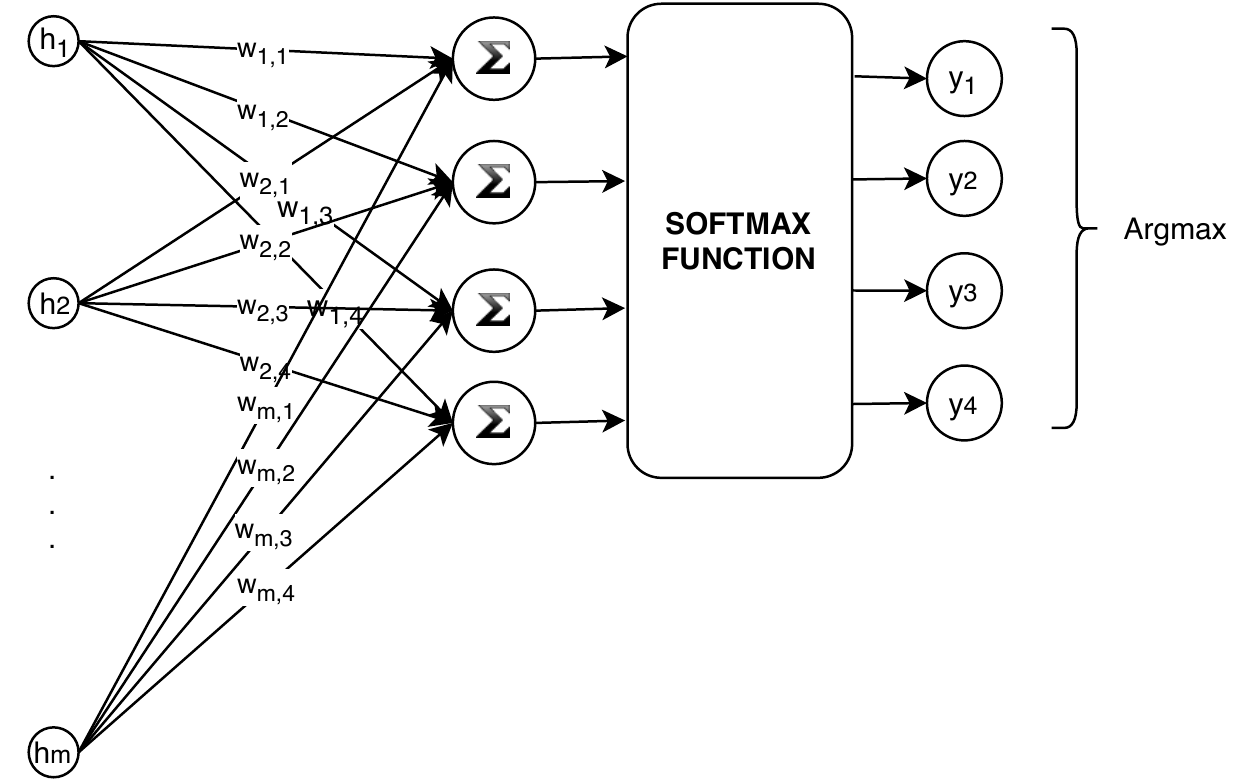}
\caption{Softmax activation on the output layer}
\label{fig:softmax}
\end{center}
\end{figure}

Softmax is used to map the non-normalized output of a network to a probability distribution over predicted output classes with probabilities between [0,1]. By assigning three independent softmax activation functions, a generalization of the binary form of logistic regression, on the output layer of the neural network for categorical target variables, the outputs can be interpreted as posterior probabilities to produce the scores for the individual tasks. 
The input to each softmax function is the result of \textit{K} distinct linear functions, \textit{K} = 4 in our architecture as shown in the Fig. \ref{fig:softmax}. 

The predicted probability for the \textit{i}'th class given a sample vector \textbf{h} and a weighting matrix \textbf{w} is:

\begin{displaymath}
 P(y=i \textbar \textbf{h})= \frac{e^{\textbf{h}\textsuperscript{T}\textbf{w}\textsubscript{i}}}{\sum_{k=1}^{K} e^{\textbf{h}\textsuperscript{T}\textbf{w}\textsubscript{k}}} 
 \end{displaymath}

This can be seen as the composition of \textit{K} linear functions \textbf{h} $\rightarrow$ \textbf{h}\textsuperscript{T}\textbf{w}\textsubscript{1}, ... , \textbf{h} $\rightarrow$ \textbf{h}\textsuperscript{T}\textbf{w}\textsubscript{K} and the softmax function (where \textbf{h}\textsuperscript{T}\textbf{w} denotes the inner product of \textbf{h}, the third hidden layer with m=20, and the weight matrix \textbf{w}). The operation is equivalent to transforming the original highly-dimensional input to vectors in a \textit{K}-dimensional space.  The standard exponential function is applied to each element \textbf{h}\textsuperscript{T}\textbf{w}\textsubscript{i} of the input vector \textbf{h}  and the values are normalized by dividing with the sum of all the exponentials; this normalization ensures that the sum of the components of the output vector \textbf{y} is 1.

Training is performed equally for all tasks using backpropagation. The function to be optimized is the mean squared error between network outputs and targets. The model learns the best parameters for predicting $\widehat{Y}$ that minimizes the loss function,

\begin{displaymath}
 L(Y,\widehat{Y})= \frac{1}{2}(Y-\widehat{Y})^{2} 
 \end{displaymath}

Models require the availability of a dataset of training and ground-truth annotations for classification. The model's accuracy strongly depends on the amount of training data and the variation within it. We present in the next section the simulation outline that help create the synthetic dataset and we provide results.

\section{SIMULATION AND RESULTS}

Economic issues and lack of large scale deployment make simulation the main choice in the validation of vehicular ad hoc networks. The realism of the simulation is thus a paramount aspect. Our experiments utilize a validated real-world traffic dataset of the City of Cologne, Travel and Activity PAtterns Simulation (TAPAS) Cologne scenario, assumed to be one of the largest traffic simulation data set \cite{uppoor2011large}. 

\subsection{Simulation outline}

TAPAS is a vehicular mobility dataset that covers the traffic road network of the city of Cologne. Different data sources and simulation tools are brought together to cover all of the specific aspects required for a proper characterization of the vehicular traffic of the dataset:

- The street layout of the Cologne urban area is obtained from the OpenStreetMap (OSM) database. The street layout shows how the area is subdivided into different segments;

- The traffic demand information on the macroscopic traffic flows across the Cologne urban area (i.e., the O/D matrix) is derived through the Travel and Activity PAtterns Simulation (TAPAS) methodology. The traffic demand is the input of the traffic assignment algorithm;

- The traffic assignment of the vehicular flows described by the TAPASCologne O/D matrix over the road topology is performed by means of Gawron’s dynamic user assignment algorithm. This results in the generation of vehicle trajectories.

- Finally, the street layout, traffic demand and traffic assignement are provided as input to a simulator. The microscopic mobility of vehicles is simulated with the Simulation of Urban Mobility (SUMO) software \cite{krajzewicz2002sumo};

The synthetic trace of the car traffic in a the city of Cologne covers a region of 400 square kilometers for a period of 24 hours in a typical working day, and comprises more than 700.000 individual car trips. We loaded the synthetic trace file as input in SUMO for the 6-8am peak hours. We call the scenario obtained, the base scenario. We further zoomed on a subsection of the region covered, precisely around a target segment where it was relevant to make short-term flow prediction.

On the other hand, SUMO enables generation of trace files that are necessary for the simulation in the network simulator ns-2 of communication between connected vehicles \cite{rehmaninetwork}. The default communication parameters in ns-2 are adjusted for DSRC (Dedicated Short Range Communications) simulations. Particularly, for the 802.11p MAC and PHY level parameters, we assume standard transmission range of the protocol, which is 300 meters. BEACON messages are exchanged every 0.1 seconds. 

We modify the base scenario to simulate atypical traffic conditions such as weather, incident, workzone, special event and recurrent congestion. In the presence of the atypical traffic conditions, vehicles are forced by the simulator to consider the speed of the leading vehicle and adapt theirs ensuing such things as road and weather conditions, traffic volumes, adjacent obstructions and distractions. The presence of the events on the base scenario thus modifies the traffic flow. This is due to the fine microscopic modelling permitted by SUMO enabling each vehicle to calculate its speed and lane choice mostly using discrete time steps of one second. We describe below the extended scenarios of atypical traffic conditions simulated using SUMO:

\underline{Scenarios of incidents and workzones:}

By incidents, we consider emergencies, accidents, vehicular breakdowns, a situation that is unexpected but has an impact on the traffic flow variables. Workzones have different characteristics and represent situations that can be planned in advance. Often times they occupy the road segment a longer period of time than incidents because incidents are undesired and should be cleared as fast as possible. Particularly, both events physically block one or multiple lanes of the road segment causing a problematic spot.

To better assess the impact of different variable combinations of the characteristics of incidents on the traffic flow, on a single segment, we simulate incidents at the beginning, middle and end of a lane. We also simulate incidents on different lanes of a segment, for a long and short duration. We insert in the simulation two vehicles in experiments where we simulate an incident on a single lane. On a three lane segment and in experiments where we block two of the three lanes, we insert three vehicles.

We simulate incidents for a long or short duration, never exceeding a duration of one hour. We then suppose that if the duration of the problematic spot is above one hour, we assume it to be a good indicator of a workzone. 
 
\underline{Extended Scenario of bad weather:}  

Inclement weather leads to slippery roads or reduced vision. This behaviour is represented in the car-following model by decreases in the speed of vehicles and defensive driver behaviour. 

\underline{Extended Scenario of a special event:} 
Regarding special events (sport games, concerts, religious activities, political demonstrations), they lead people to travel towards the same destination in a very limited time interval. To simulate this behaviour, we generate trips to a particular destination edge of the city of Cologne. A trip is a vehicle movement from one place to another defined by the starting segment, the destination segment, and the departure time. A route is an expanded trip that means that a route definition contains not only the first and the last segment, but all segments the vehicle will pass along its trajectory.

We fix the destination segment where the special event is located and generate random starting segments (departures) and routes to simulate different inbound traffic of diverse demand categories of people going to the event. A Poisson process is used for random timings for trips and departures will occur individually, stochastically independent to all the others in the road network, at random moments. We vary the parameters to populate the dataset of different scenarios of special events. 

When vehicles on the congested segment experiencing congestion detect that the observed travel time is excessive, the observable trajectory characteristics and the results of the local traffic evaluation are collected. We generate urban mobility traces from the extended scenarios for usage in ns-2 in order to carry on the cooperative process. Data of independent vehicles passing on the target road segment are collected. Characteristics are extracted from several scenarios, experiments and vehicles respectively and put into supervised feature vectors. We construct a synthetic training dataset. 

All inputs of the prediction model are real numbers except for the traffic events happening on the segments. Traffic events consist of one of the six causes: \big \langle accident, workzone, weather condition, recurrent, special event or no events \big \rangle. This order of the causes is important because the values representing events are given to the model via an encoding. The encoding we use is a one-hot vector of six values either being 0 or 1 in the presence of the event in the previous order. For instance, in presence of an accident, the one-hot vector is \big \langle 100000 \big \rangle,  and in presence of a workzone, the vector is \big \langle 010000 \big \rangle.

Once we have obtained the features, we map the prediction variables to discrete classes by dividing the range of values into sub-ranges. We train the STP model. 

\subsection{Results}

\begin{figure}[H]
\begin{center}
\includegraphics[width=1\linewidth,height=9cm]{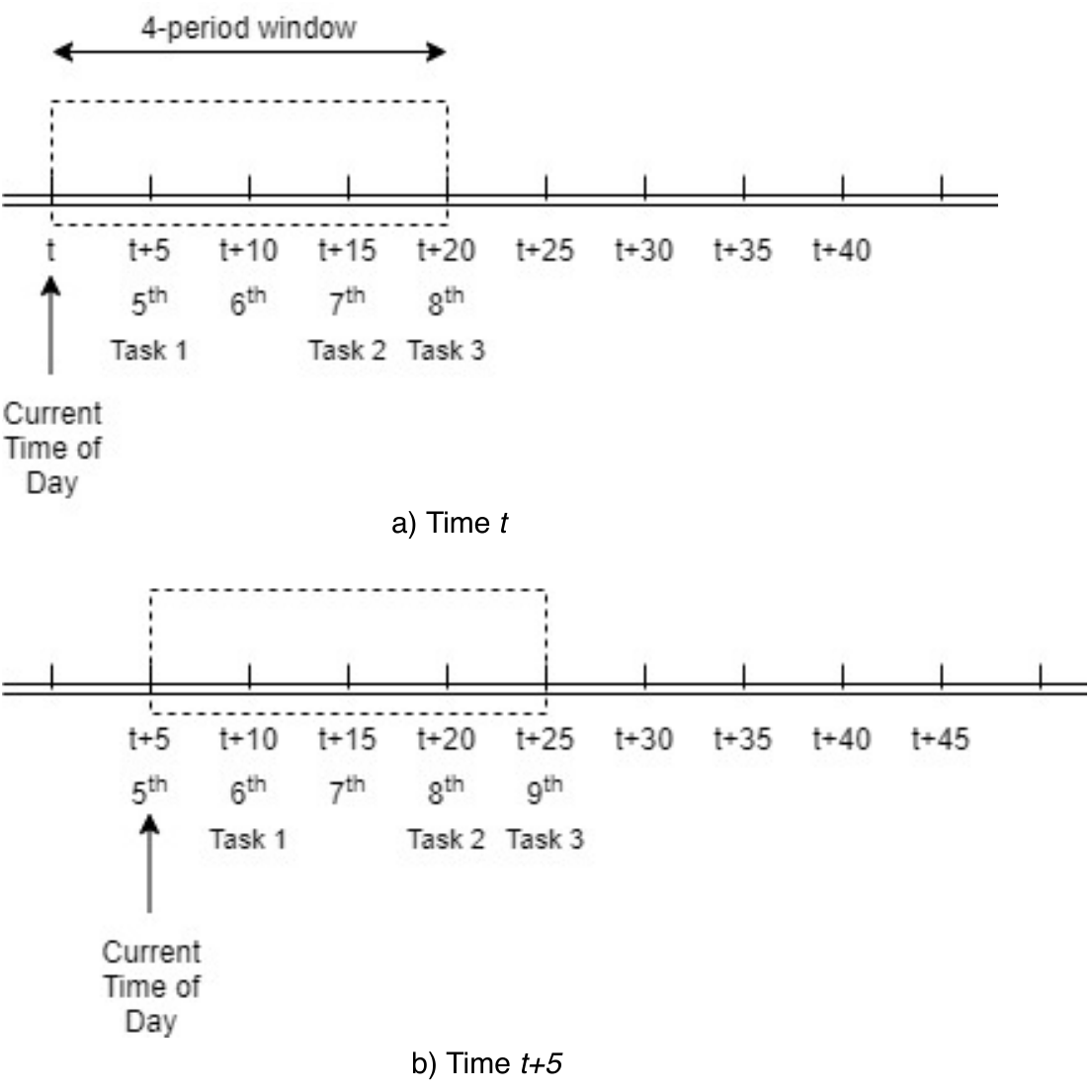}
\caption{4-period window and corresponding labels of the input vector}
\label{fig:window}
\end{center}
\end{figure}

In our study, flows from 6:00 am to 8:00 am are collected on a target road segment under different scenarios. Every period of five minutes, an input in the dataset corresponding to a feature vector is created. As per the formulation of our supervised learning model, each input vector must be tagged with the future flow values. Thus, every input must have three labels corresponding to the three tasks of prediction of the classifier happening 5, 15 and 20 minutes later. To create the input at time \textit{t}, the simulation must have ran until \textit{t+20}, corresponding to a 4-period window as per Fig. \ref{fig:window}a). For a current time of day corresponding to \textit{t}, flow at \textit{t+5} corresponds to the \textit{5th} data point and represents the first label of the input. Since the second task at time \textit{t} is to predict flow at \textit{t+15}, this represents the \textit{7th} data point in the figure and represents the second label of the input. The last task at time \textit{t} is to predict flow at \textit{t+20}, and this represents the \textit{8th} data point in the figure. 

Fig. \ref{fig:window}b) shows the window for the following period in order to create the input at time \textit{t+5}. The simulation must have ran until \textit{t+25}. The \textit{6th} data point represents now the first label of the input corresponding to the first task. The second label corresponds to the \textit{8th} data point occurring at \textit{t+20} and the third label becomes the \textit{9th} data point captured at \textit{t+25} of the simulation. This process continues until the last observation.

\begin{figure*}[!t]
\begin{center}
\includegraphics[width=1\linewidth,height=4.5cm]{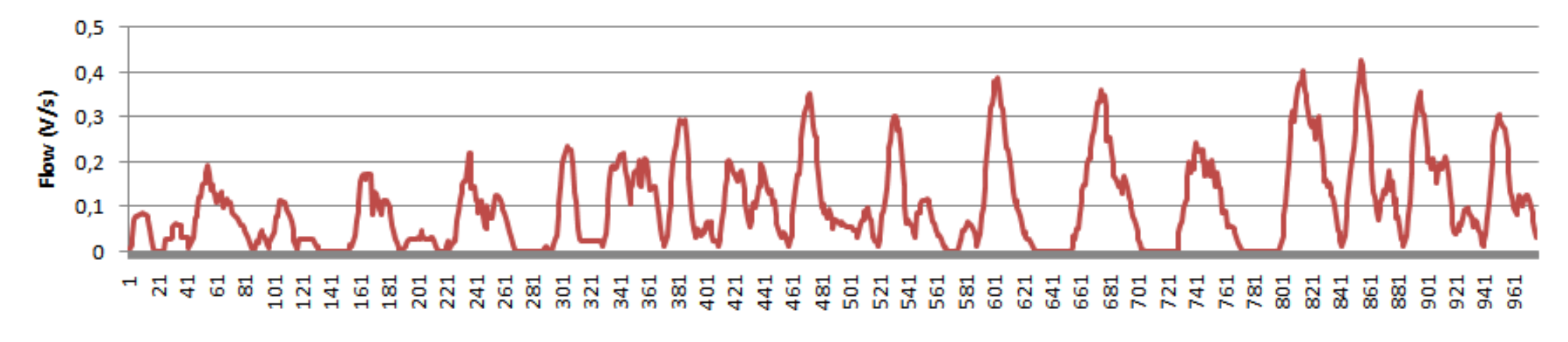}
\caption{Traffic flow on a signalised road segment - no events}
\label{fig:flowrecurrent16945}
\end{center}
\end{figure*}

\begin{figure*}[!t]
\begin{center}
\includegraphics[width=1\linewidth,height=5cm]{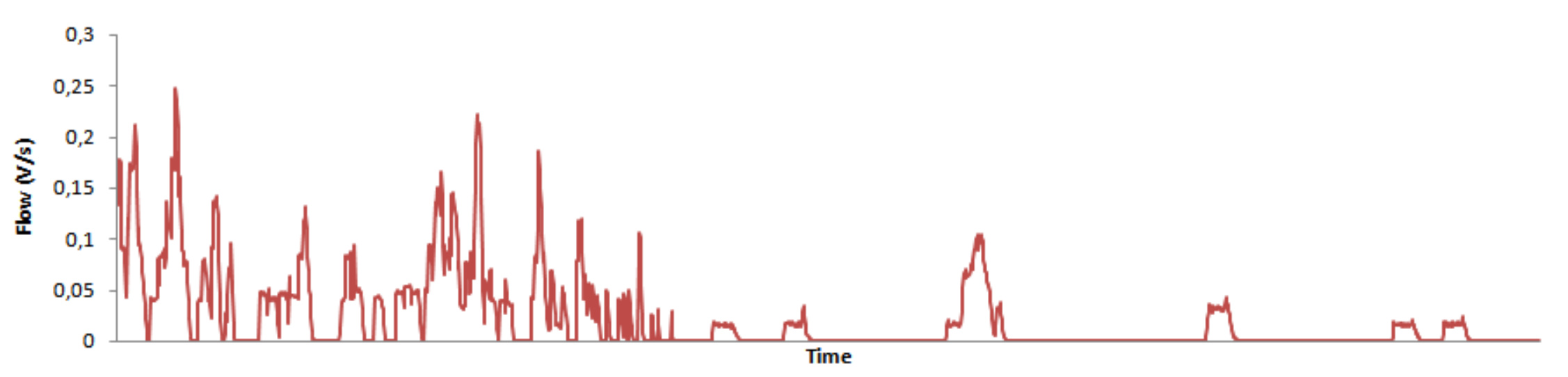}
\caption{Traffic flow on a signalised road segment - Event: incident}
\label{fig:flowincident13444}
\end{center}
\end{figure*}

We plot in Fig. \ref{fig:flowrecurrent16945} the traffic flow values for a signalised road segment in the urban network. From this figure, we can sense that vehicles stop at the light and then another wave starts periodically. This behaviour of traffic flow is different from that occurring on a freeway. In the figure, no congestion and no events are present on the road segment. In Fig. \ref{fig:flowincident13444} however, we show the profile of traffic flow values in the advent of an incident occurring on the urban road segment. We notice as congestion installs, how the flow values stay very low because density is high. We demonstrate how our model accurately predicts future flows in presence of any cause of congestion. 

To make the proposed framework tractable, we compare the performance of our MTL-CV model with various prediction models. Firstly, since our multitask learning model is built on MLP network, it is worth to compare and investigate how much improvement we could achieve beyond the baseline ANN classifier. In fact, we experimented with three net architectures. ANN is a standard net that learns the task of short term traffic flow prediction 15 minutes later. MTL\textsubscript{a} is a net that learns two tasks of prediction of traffic flow with the first task being prediction of flow 5 minutes later, that is before the target time and the second task is the main task of prediction of short term traffic flow 15 minutes ahead. Finally, MTL\textsubscript{b} is a net that learns two tasks of prediction of traffic flow with the first task being the main task of prediction of short term traffic flow 15 minutes later and the second task is prediction of flow 20 minutes later.

Also, to measure the predictive power of the proposed MTL-CV model, we compared it with the performance of the state-of-the-art ARIMA time series approach and with a baseline classifier, Random Forest (RF), implemented in Weka \cite{Hall:2009:WDM:1656274.1656278}. MTL-CV, MTL\textsubscript{a} and MTL\textsubscript{b} are implemented using Torch 5 package. Specifically, when evaluating the performance of our model, we use root-mean-square error, RMSE.  

\begin{displaymath}
RMSE= \sqrt{\frac{1}{T}\sum_{t=1}^{T}(Yt-\widehat{Yt})^{2}}
\end{displaymath}

We feed ARIMA the original traffic flow data. For the Auto Regressive part we do iterations of 1 to 10, followed by 1 to 10 iterations for Moving Average. We use the model to predict the next value. We iterate the forecasting procedure five times by using the predicted flow as previously observed flow. For RF and the neural networks however, the data is not only traffic flows but also the other features proposed in this paper. The dataset is then divided in training, validation and test sets. For the one-hidden-layer ANN, units are varied between [5, 150] in steps of 5, the number of epochs is varied between [25, 250] in steps of 25. They are set through cross-validation. We added dropout between all the layers of the network to improve generalization. The best architecture that we obtained for ANN has 90 hidden units, and the number of epochs is 150.

On the other hand, for the deep networks, more hyper-parameters have to be considered, nodes in each layer, the size and the epochs. The evaluation is based on five folds cross-validation with 20 randomly repeated cross-validation runs on the training set to obtain average performance scores for comparisons. Backpropagation is done on all outputs. To avoid overfitting, we did not try very deep MLP architecture. In the case of MTL-CV, MTL\textsubscript{a} and MTL\textsubscript{b}, we tested two to five layers, [5, 150] in steps of 5 nodes in each layer and 50 to 300 epochs in the steps of 50. The best architecture that we obtained for MTL-CV has 20, 40, and 20 hidden units in the first, second, and third hidden layers. The best number of epochs is 100. 

Table~\ref{weather} shows results of MTL-CV in comparison with the time series, baselines and MTL\textsubscript{a} and MTL\textsubscript{b} using RMSE. The scores are averaged from 20 randomly repeated 5-folds cross-validation runs.

\begin{table*}[!t]
\renewcommand{\arraystretch}{2.8}
\caption{Performance comparison of MTL-CV with the time series, baselines (RF, ANN) and MTL\textsubscript{a} and MTL\textsubscript{b} using RMSE. }
\label{weather}
\centering
\begin{tabular}{llllllll} 
\hline 
Task & ARIMA & RF & ANN & MTL\textsubscript{a} & MTL\textsubscript{b} & MTL-CV \\
\hline
5-min Traffic Flow prediction & - & - & - & 0.042 & - & 0.056\\
\hline
\hline
15-min Traffic Flow prediction & 0.255 & 0.122 & 0.113& 0.073 & 0.085& 0.052\\
\hline
\hline

20-min Traffic Flow prediction & - & - & - & - & 0.094 & 0.108\\

\end{tabular}
\end{table*}

ANN model makes comparable performance to the state-of-the-art RF model. But RF and ANN achieve better performance than the ARIMA time series, with error values of 0.122 and 0.113 respectively.  This is expected because of the added features in the input vector. Also, deep networks perform better than baseline ANN because deep networks learn sub-features in the different layers to better characterise the output flow. Consequently, in all scenarios of traffic congestion due to different event, deep networks track better the sudden flow changes and their pattern. Particularly, results indicate that multi-tasking improve the performance compared to single task learning with ANN. Task 2 in MTL\textsubscript{a} and task 1 in MTL\textsubscript{b} try to capture the information contained in the training signals of other tasks drawn from the same domain. The tasks in these models exploit the joint input. If the tasks can share what they learn, the model performs better when it learns them together than in isolation. The difference between MTL\textsubscript{a} and MTL\textsubscript{b} is in the training phase. Because of the joint representation, MTL\textsubscript{a} is 0.04 better than ANN on the test set result and MTL\textsubscript{b} is about 0.028 better than ANN.

We analyze the contribution of MTL-CV to the prediction problem. We notice that some hidden units of MTL\textsubscript{a} and MTL\textsubscript{b} became specialized for just one or a few tasks. Task 2 in MTL\textsubscript{a} and Task 1 in MTL\textsubscript{b} need to compute the same subfeatures. If Task 1 from MTL\textsubscript{a} and Task 2 from MTL\textsubscript{b} are used as extra outputs in MTL-CV, this signify that they must be learned; it will bias the shared hidden layer to learn the input features better, and this will help the MTL-CV net better learn to predict outputs. This confirms the importance of having highly related tasks and our idea of using MTL-CV to improve the target Task 2. MTL-CV provides the best RMSE, improving MTL\textsubscript{a} and MTL\textsubscript{b} by 0.021 and 0.033, respectively. Indeed, Tasks 1 and 3 help solving it. Generalization in neural nets improved because the net learned to better represent underlying regularities of the domain.  MTL-CV provides a benefit with time series data because predictions at different time scales often partially depend on different tasks.

On another hand, the data resolution provided by the connected vehicles technology is another reason behind the high performance of our model. Aggregation of high-resolution raw data into lower resolution levels is a common practice in short-term traffic forecasting studies. In our case, since data are exchanged between connected vehicles every 0.1 seconds, we face the opposite case where we aggregate into high-resolution. 

We conducted a sensitivity analysis on the features of the model to uncover the importance of a feature for some partial classification. We removed one feature at a time and used the filtered training set for classification. The sensitivity analysis showed that some features had no impact on the accuracy of the model, we removed them and considered them to be non-predictive attributes. We ended up with the 62 input features of the model we propose because they are the most relevant. The benefits of revealing the most relevant context dimension are manifold, including reduced cost due to context information retrieval and transmission, reduced algorithmic and computation complexity. 

Table~\ref{others} shows results of MTL-CV for the main task of 15-min Traffic Flow prediction for the special event, recurrent congestion, incident and workzone scenarios using RMSE. We carefully analysed the instances where the MTL-CV model made errors. We found that mistakes were made mostly from incident and workzone scenarios. This is due to the fact that two traffic situations may be represented by the same input features, but the output label of one of the tasks can be different. In fact, we monitor the same network region at same time for two traffic situations. In traffic situation-1, with an incident at the top of the segment, the label at \textit{t+15} of the MTL-CV model can be the same as the label at \textit{t+15} in traffic situation-2, where there is an incident on the same segment but this time at the bottom of it, however, the label \textit{t+20} is different in both situations although they have the same input. 

\begin{table*}[!t]
\renewcommand{\arraystretch}{2.8}
\caption{Performance comparison of MTL-CV for the main task of 15-min Traffic Flow prediction for the special event, recurrent congestion, incident and workzone scenarios using RMSE.}
\label{others}
\centering
\begin{tabular}{llllllll} 
\hline 
\hline 
Main task & ARIMA & RF & ANN & MTL\textsubscript{a} & MTL\textsubscript{b} & MTL-CV \\
\hline
\hline 
Special Event & 0.252 & 0.112 & 0.130 & 0.104 & 0.119 & 0.098\\

Recurrent & 0.160 & 0.135 & 0.145& 0.139 & 0.155 & 0.103\\

Incident & 0.308 & 0.343 & 0.317 & 0.276 & 0.205 & 0.198\\
Workzone & 0.367 & 0.301 & 0.298 & 0.264 & 0.223 & 0.215\\

\end{tabular}
\end{table*}

In this context, the incident happening on the street had the same effect on flow in a short period but later on it cleared better. The position of the incident on the road segment had an impact on flow later on in time because maybe meanwhile, the incident cleared better. We analysed the situation and found lots of vectors in the dataset that behave like this in the incident and workzone scenarios only. One way to solve this problem might be by changing the network architecture. In fact, the MTL-CV net presented in this article use fully connected hidden layer shared equally by all tasks. A more complex architecture would be a small private hidden layer for the main task, and a larger hidden layer shared by both the main task and extra tasks. With no principled ways to determine what architecture is best for each problem, only testing will tell.

Finally, in seemingly normal traffic conditions, our proposal helps to understand how effects that could be classified as negligible or simply cannot be handled with a traditional, human in the loop approach, may have dramatic impacts on the road network. This is due to the fact that the inference of future traffic flow values in our framework is done with regards to information only the vehicles themselves know of and no human could have possibly been able to assess. We realized that the scenarios falling under this category of seemingly normal traffic condition must have been due to any of the causes of non-recurrent congestion. The only way to obtain this traffic behaviour via simulation was by conducting slight variation in the experiments of the non-recurrent congestion scenarios. Results for these scenarios, classified according to the underlying non-recurrent congestion simulated, are averaged and represented in Table~\ref{others}.  

\subsection{Impact of the communication network}

We study the impact of the communication network to provide a more effective and efficient evaluation of the proposed MTL-CV model in terms of resiliency. We conduct experiments where we assess the performance of the model for the task of 15-min traffic flow prediction on a target road segment in a one at a time analysis of the parameters; we study only one parameter while the others remain fixed. We study the impact of two parameters: the travel time index (TTindex) and the flow estimated by the CV. 

The simulator SUMO will provide the real values for the parameters to be estimated and flows to be predicted. This will provide the ground truth. On the other hand, in the simulation of the wireless technology where radio channel effects, packet losses and delay alter packet flow to mimic a realistic behaviour, we are able to extract the same parameters. However, we call them estimated parameters since their accuracy is slightly inferior. We provide the estimated parameters from each simulation as input to the MTL-CV to study the impact on the prediction of flow. We then conduct the same simulation on different segments, compute the Root Mean Square Error (RMSE) and average the total. We maintained the same scenario of congestion caused by an incident on a segment and traffic flow to be predicted on a neighboring signalized road segment. 

\begin{figure*}[!t]
\begin{center}
\includegraphics[width=1\linewidth,height=9cm]{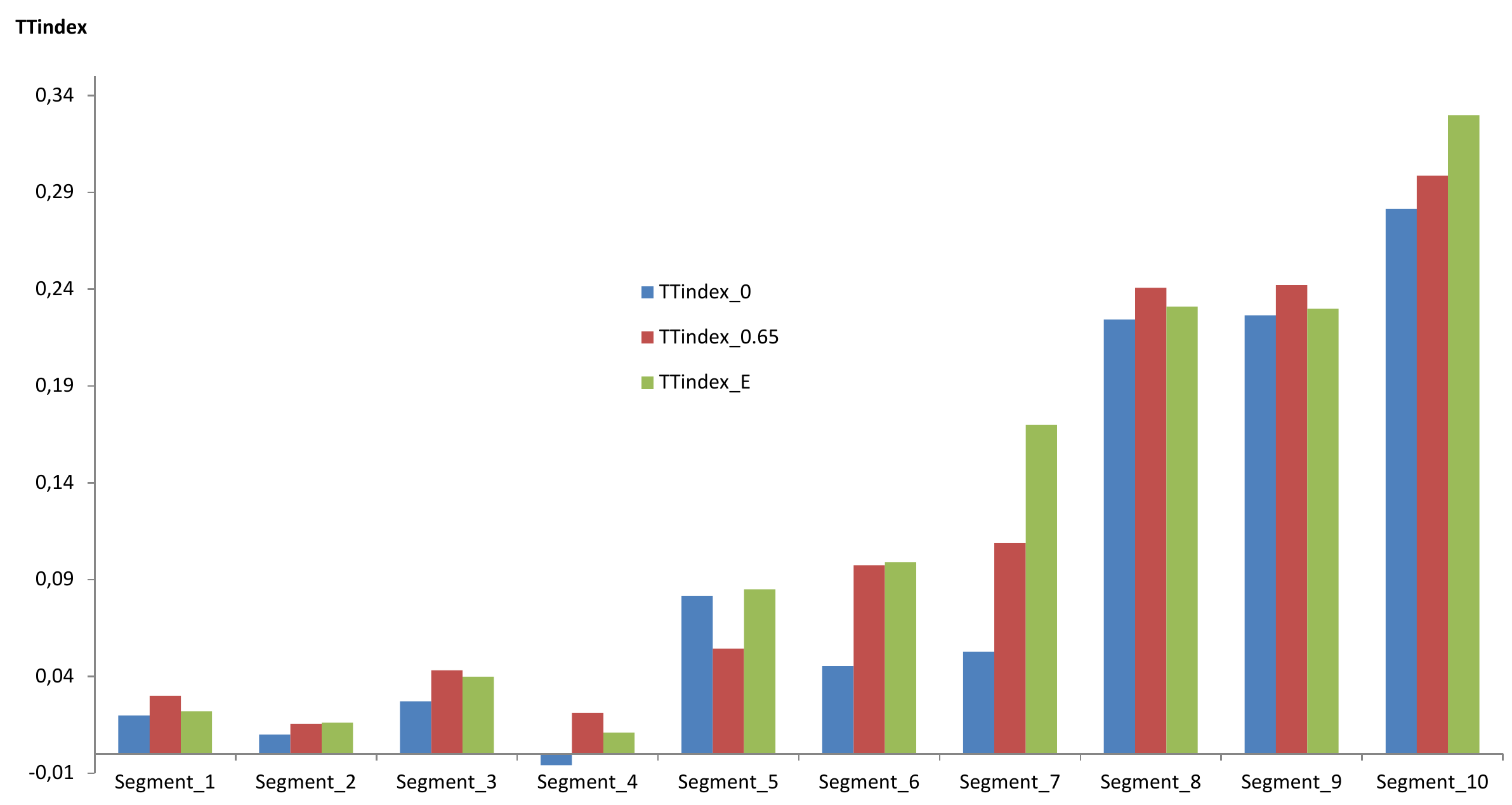}
\caption{TTindex values of a vehicle along its trajectory under different conditions - TTindex\textunderscore0 : Worst case when \textalpha=0, TTindex\textunderscore0.65 : Ground-truth provided by SUMO with \textalpha=0.65 , TTindex\textunderscore E : Estimated TTindex values exchanged via the wireless network. }
\label{fig:EstimatedTTindex}
\end{center}
\end{figure*}

\subsubsection{Impact of the travel time index}

The travel time index, TTindex, is computed on a per-segment basis. In Fig.\ref{fig:EstimatedTTindex}, we compare the TTindex values of a vehicle for the last 10 segments of its trajectory under different conditions. Each vehicle computes travel time on each segment of its trajectory in order to compute a TTindex of their own as per equation (1). At this stage, there is no exchange via the wireless network of any information between the vehicles and the TTindex\textunderscore0 computed represents the worst case in the figure where \textalpha=0. A CV may collect travel time indexes of others by cooperation in order to compute the average. The average is represented in the Figure and TTindex\textunderscore0.65 is calculated using \textalpha=0.65 in Equation (2) of the paper. We used this computation in our model and label the values as the ground truth. The estimated TTindex values, TTindex\textunderscore E in the figure represent the values assessed via the wireless network as they are based on information contained in exchanged and received packets.  

We notice that compared to the ground truth, the estimation of TTindex has a RMSE of 0.0246 while TTindex\textunderscore0 has an error of 0.0292 for the portion of trajectory represented in the figure. The RSU collects the TTindex values, computes the average and feeds the average to the MTL-CV model to predict the flow. The model predicts future flows in the advent of an incident occurring on a neighboring segment when estimated TTindex values are used. In comparison to the ground truth, we observed an error increase of 0.119 in terms of RMSE.

\subsubsection{Impact of flow values estimated by the CV}

\begin{figure}[h]
\begin{center}
\includegraphics[width=1\linewidth,height=6.5cm]{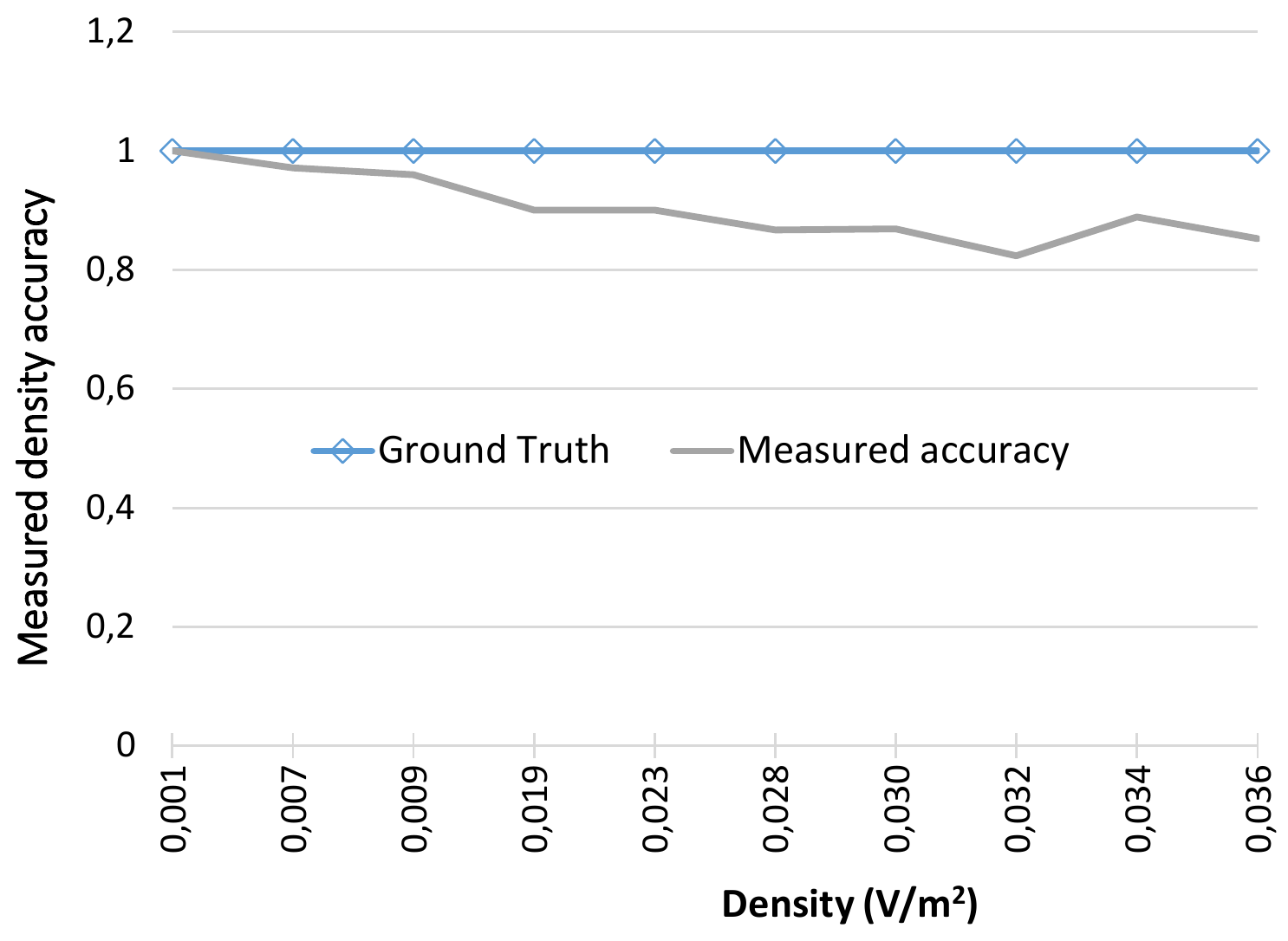}
\caption{Variation of the measured density accuracy on a road segment.}
\label{fig:DensityAccuracy}
\end{center}
\end{figure}

We evaluate the effect of the flow estimated by the CV on the performance of our model. We illustrate graphically in Fig.\ref{fig:DensityAccuracy} the measured density accuracy. We present the density because each connected vehicle uses this parameter in its estimation of flow. Flow is the number of vehicles detected per period of time. We compute the ratio between the number of vehicles detected by the CV on the segment and the number of existing vehicles. This ratio represents the measured density accuracy. In the figure, the traffic density on the segment varies from 0.0001 to 0.036 a vehicle per square meter. The figure shows that the CV are not able to detect the entire number of vehicles on the investigated road segment and that the accuracy decreases in conjunction with increases in road density. This is caused mainly by lost messages. The higher the density, the greater the number of sent messages and the communication network becomes congested.

If a vehicle misses some messages of traveling vehicles on its current road segment, its evaluation of traffic flow will not be accurate. Compared to the short-term traffic flow predictions done by the MTL-CV model with ground truth flow values, the performance of the model is lower for flow values estimated by the CV, with an error of 0.197. This shows the impact of the very dynamic vehicle’s mobility in urban environment. The vehicle’s overall knowledge about the real-time flows is slightly altered because of the short-term changes in vehicle’s motion. Also, in the case where all vehicles broadcast their messages at the same time, most of these messages will collide with each other; this will occur despite the use of CSMA method to control the communication channels. 

The challenges due to the communication network in terms of channel conditions, packet losses, collisions and delay have an impact on the traffic flow prediction. Numerous research studies have investigated the adaptive message transmission in a vehicular environment to solve many communication network problems \cite{thaina2011study}. In general, any of the solutions can be used to reliably disseminate the basic traffic data over the network. Also, a protocol at the communication level that systematically goes through the steps that each vehicle should follow to evaluate the parameter, may help the value estimated be more accurate in presence of the collision occurrences over the communications network.

\section{CONCLUSION}

We proposed a Short-term Traffic flow Prediction (STP) framework for urban road networks so that transportation authorities take early actions to control the flow and prevent congestion.  The framework is semi-centralized because on one hand, connected vehicles (CV) collect and propagate data via the ad hoc networks they form between each other along a route. A road side unit (RSU) is installed on a target road segment and collects data for a period of time to get a clearer picture of the traffic on the road. To cope with the fact that current research on traffic prediction mainly focuses on data traffic history and neglects other conditions affecting traffic, in this paper, we showed how CV technology allow advanced modelling by integrating into the forecasting of flow, the impact of the various events that CV realistically encountered on segments along their trajectory. We solved the STP problem with a neural network. Our deep architecture in a multitask learning setting (MTL-CV) showed its advantage for a complex urban transportation system. Our experiments on a synthetic dataset show that the results of our approach significantly outperforms state-of-the-art ARIMA time series and baseline classifiers, with an average root-mean-square error (RMSE) of 0.05. Compared to single task learning with Artificial Neural Network (ANN), ANN had a lower performance (0.113 for RMSE) than MTL-CV. The analysis of the impact of overestimating the performance of the wireless network showed that an estimated travel time index had an increase in error of 0.119 on the traffic flow prediction. The performance of the model is even lower for flow values estimated by the CV, with an increase in error of 0.197 on the prediction.

Adding heterogeneity can be explored in future work as there could be various sources that could bring heterogeneity in vehicular behavior. One of them is to reproduce synthetically safe distances. We can also simulate the fact that drivers may take different paths in the event of an incident because in reality, drivers may be alerted about an event and change their trajectory, thus changing vehicular flows. Since realistic datasets of the vehicular dynamics of the same region under different traffic scenarios (weather condition, special events, ...) are not available, studying the heterogeneity help to understand the impact in more realistic environments and to provide more efficient evaluation of the model. 

\bibliographystyle{IEEEtran}
\bibliography{IEEEabrv,test}

\begin{IEEEbiography}
    [{\includegraphics[width=1in,height=1.25in,clip,keepaspectratio]{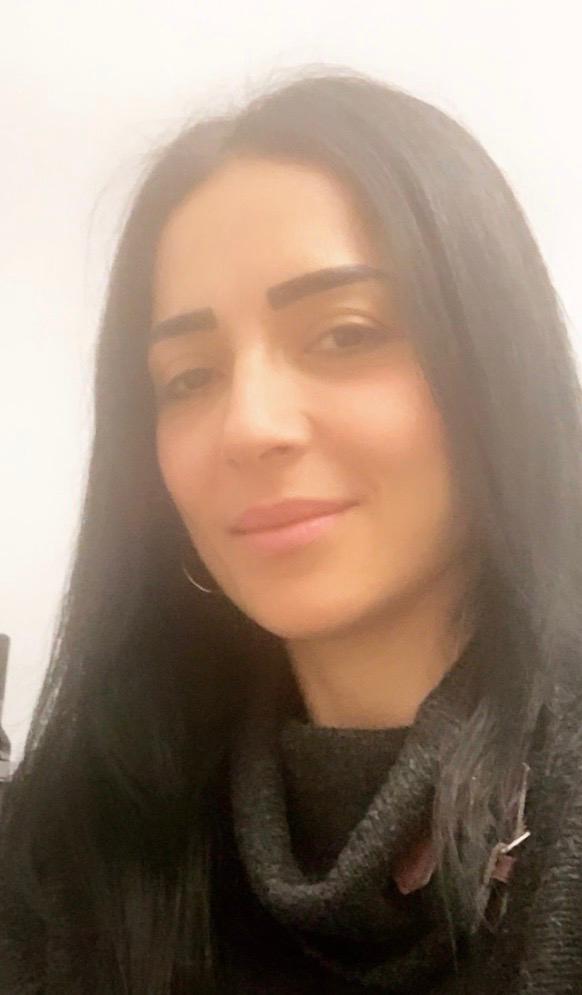}}]{Ranwa Al Mallah}
received her Ph.D. degree in Computer Science from Polytechnique Montreal, Canada. Her current research interest includes cybersecurity of Intelligent Transportation Systems, Traffic Efficiency and Safety Applications.
\end{IEEEbiography}

\begin{IEEEbiography}
    [{\includegraphics[width=1in,height=1.25in,clip,keepaspectratio]{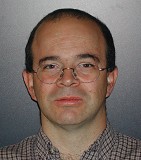}}]{Alejandro Quintero}
is currently a full professor at the Department of Computer Engineering of  Polytechnique de Montreal, Canada. Dr. Quintero is the coauthor of two book, as well as over 100 other technical publications including journal and proceedings papers.
\end{IEEEbiography}

\begin{IEEEbiography}
    [{\includegraphics[width=1in,height=1.25in,clip,keepaspectratio]{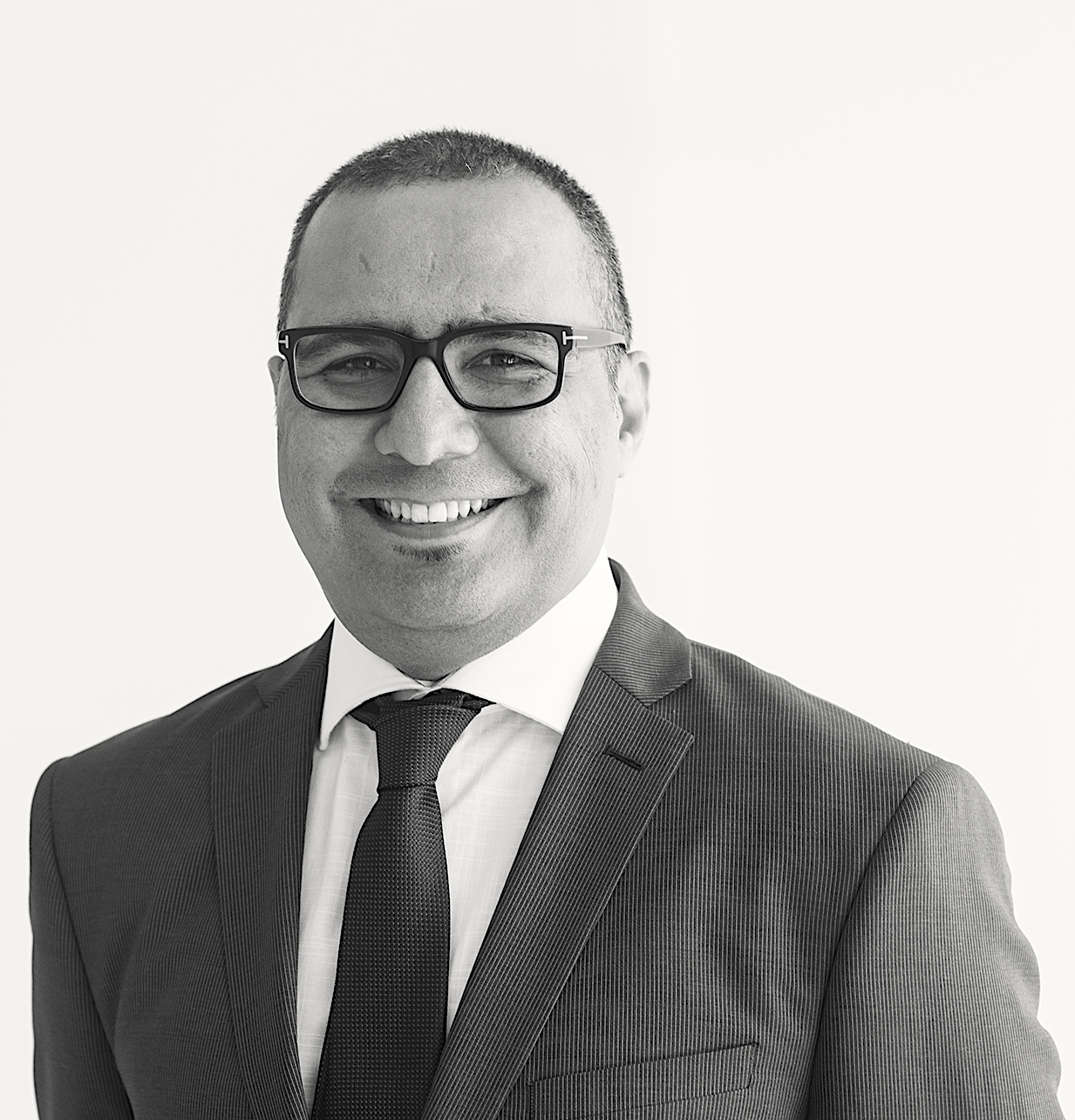}}]{Bilal Farooq}
received his Ph.D. degree from University of Toronto, Canada in 2011. His current research goal is to develop multidisciplinary and highly-intelligent solutions for sustainable planning, design, and operations of urban infrastructure systems.
\end{IEEEbiography}

\end{document}